\documentclass[pra,aps,twocolumn,nopacs,superscriptaddress,nofootinbib]{revtex4}

\usepackage{amsmath}  \usepackage{amssymb}  \usepackage{amsfonts}  \usepackage{bm}  \usepackage{bbm}   \usepackage{bbold}   \usepackage{braket}  \usepackage{color}  \usepackage{comment}  \usepackage{dcolumn}  \usepackage{enumerate}  \usepackage{epsfig}  \usepackage{gensymb}  \usepackage{graphicx}  \usepackage{indentfirst}  \usepackage{lmodern}  \usepackage{mathrsfs}  \usepackage{mathtools}  \usepackage{psfrag}  \usepackage{pst-all}  \usepackage{soul}  \usepackage{xcolor}
\usepackage{float} %---APS---SI---arxiv
\usepackage[colorlinks,linkcolor=blue,citecolor=blue,urlcolor=blue,hyperindex,driverfallback=dvipdfm]{hyperref}  \usepackage[T1]{fontenc} %---APS---ACS---SI---arxiv

\def\ii{{\rm i}} \def\ee{{\rm e}}

\def\me{m_{\rm e}}
\def\Eb{{\bf E}}

 \def\Ab{{\bf A}}
\def\Hint{\mathcal{H}_{\rm int}}

\def\NA{{\rm NA}}

 \def\NA{\rm NA}

\begin{document} %---APS---SI---arxiv

% =========================================================
% --- title, affiliations, abstract -----------------------
% =========================================================
\title{Free-Space Optical Modulation of Free Electrons in the Continuous-Wave Regime}

\author{Cruz~I.~Velasco}
\affiliation{ICFO-Institut de Ciencies Fotoniques, The Barcelona Institute of Science and Technology, 08860 Castelldefels (Barcelona), Spain}
\author{F.~Javier~Garc\'{\i}a~de~Abajo}
\email{javier.garciadeabajo@nanophotonics.es}
\affiliation{ICFO-Institut de Ciencies Fotoniques, The Barcelona Institute of Science and Technology, 08860 Castelldefels (Barcelona), Spain}
\affiliation{ICREA-Instituci\'o Catalana de Recerca i Estudis Avan\c{c}ats, Passeig Llu\'{\i}s Companys 23, 08010 Barcelona, Spain}

% --- abstract --------------------------------------------
\begin{abstract}
The coherent interaction between free electrons and optical fields can produce free-electron compression and push the temporal resolution of ultrafast electron microscopy to the attosecond regime. However, a large electron--light interaction is required to attain a strong compression, generally necessitating short light and electron pulses combined with optical scattering at nanostructures. Here, we theoretically investigate an alternative configuration based on stimulated Compton scattering, whereby two counterpropagating Gaussian light beams induce energy jumps in a colinear electron beam by multiples of their photon-energy difference. Strong recoil effects are produced by extending the electron--light interaction over millimetric distances, enabling a dramatic increase in temporal compression and substantially reshaping the electron spectra for affordable laser powers. Beyond its fundamental interest, our work introduces a practical scheme to achieve a large temporal compression of continuous electron beams without involving optical scattering by material structures.
\end{abstract}
\date{\today}

\maketitle %---APS---OSA---SI---arxiv
\date{\today} %---APS---arxiv

% =========================================================
\section{Introduction}

The short de Broglie wavelength of free electrons ($2-39$~pm) at the energies commonly employed in electron microscopes ($1-300$~keV) is currently leveraged to reach an unsurpassed combination of spatial and energy resolution thanks to the development of aberration correctors compatible with a numerical aperture (NA) of $\lesssim1/50$ \cite{NCD04,MKM08} as well as monochromators and energy analyzers reaching a precision of a few meV \cite{KUB09,KS14,HNY18}. In addition, exposure to optical fields (amplitude $\mathcal{E}_0$, frequency $\omega$, interaction length $L$) can shape the electrons into trains of sub-femtosecond pulses \cite{BZ07,FES15}, provided the electron--light interaction parameter $\beta\sim e\mathcal{E}_0L/\hbar\omega$ \cite{paper371} reaches values close to unity. In the photon-induced near-field electron microscopy (PINEM) technique \cite{BFZ09}, $\beta>1$ is produced by resorting to ultrafast lasers to generate femtosecond photoelectron pulses that interact with the near field scattered by a material structure illuminated by synchronized femtosecond light pulses. Playing with the delay between electron and light pulses, a femtosecond \cite{PLQ15,paper282,VMC20,KDS21} (and even attosecond \cite{NKS23,paper431,BNS24,YFR21,paper415}) temporal resolution can be achieved while reaching interaction coefficients $\beta\gg1$ with $L\sim1$~$\mu$m's, $\omega$ in the visible/near-infrared domain, and $\mathcal{E}_0\sim1$~GV. Nonetheless, when a single light frequency is employed, a material structure is needed to mediate this form of stimulated inelastic electron--light scattering (SIELS), thus limiting the duration and strength of the optical field to avoid material damage. SIELS mediated by dilute gases is also possible \cite{VD1971,WHC1977}, demanding further investigation into the ability of this configuration to produce temporal electron shaping.

SIELS in free space is possible via stimulated Compton scattering by, for example, using two optical fields with different frequencies $\omega_1$ and $\omega_2$ to produce energy jumps $\hbar(\omega_1-\omega_2)$ in the electron \cite{KES18,KSH18,TTB23}. This effect is mediated by quadratic terms in the optical vector potential $\Ab$ within the electron-light interaction Hamiltonian, whose strength is larger than the linear terms responsible for PINEM if the field amplitude $\mathcal{E}_0$ exceeds $\sim\me v\omega/e $ (i.e., $\mathcal{E}_0>10^2~$GV/m for $\hbar\omega=1$~eV with an electron kinetic energy above 1~keV). Such large field amplitudes can be reached using pulsed electrons and lasers to produce stimulated Compton scattering \cite{KES18,KSH18,TTB23}.

Temporal electron shaping in the continuous-wave (CW) regime would be advantageous in avoiding the costly operations of single-electron counting and electron--light pulse synchronization, thus enabling higher electron-beam (e-beam) currents. A possible strategy for producing a large coupling coefficient $\beta$ under CW conditions (i.e., with field amplitudes $\mathcal{E}_0$ below the material damage threshold) consists in increasing the interaction length $L$, as proposed \cite{paper180} and recently demonstrated \cite{FHA22,HEK23} for the efficient generation of single photons by free electrons. Strong SIELS with $\beta\gg1$ has been demonstrated using CW light and electron beams by using microring resonators that can reach high levels of optical intensity, although problems with electrical charging of the cavities arise if the e-beam passes close to the surface. Undesired electron-surface collisions are avoided in free space, where strong phase imprinting on free electrons has been proposed \cite{paper368} and demonstrated \cite{MWS22} to shape the lateral electron profile using structured light pulses. Phase imprinting in the CW regime has also been demonstrated by using strong focused lasers \cite{MJD10,SAC19,ACS20}. These advances rely on quasimonochromatic light, so the electron--light interaction is elastic. Using two-color illumination instead, free space SIELS could facilitate a temporal shaping of the electron, which has been explored using pulsed lasers \cite{KSH18,TTB23,ET23}, but its extension to the CW regime remains a challenge.

% --- Figure 1 --------------------------------------------
\begin{figure}
\centering\includegraphics[width=0.4\textwidth]{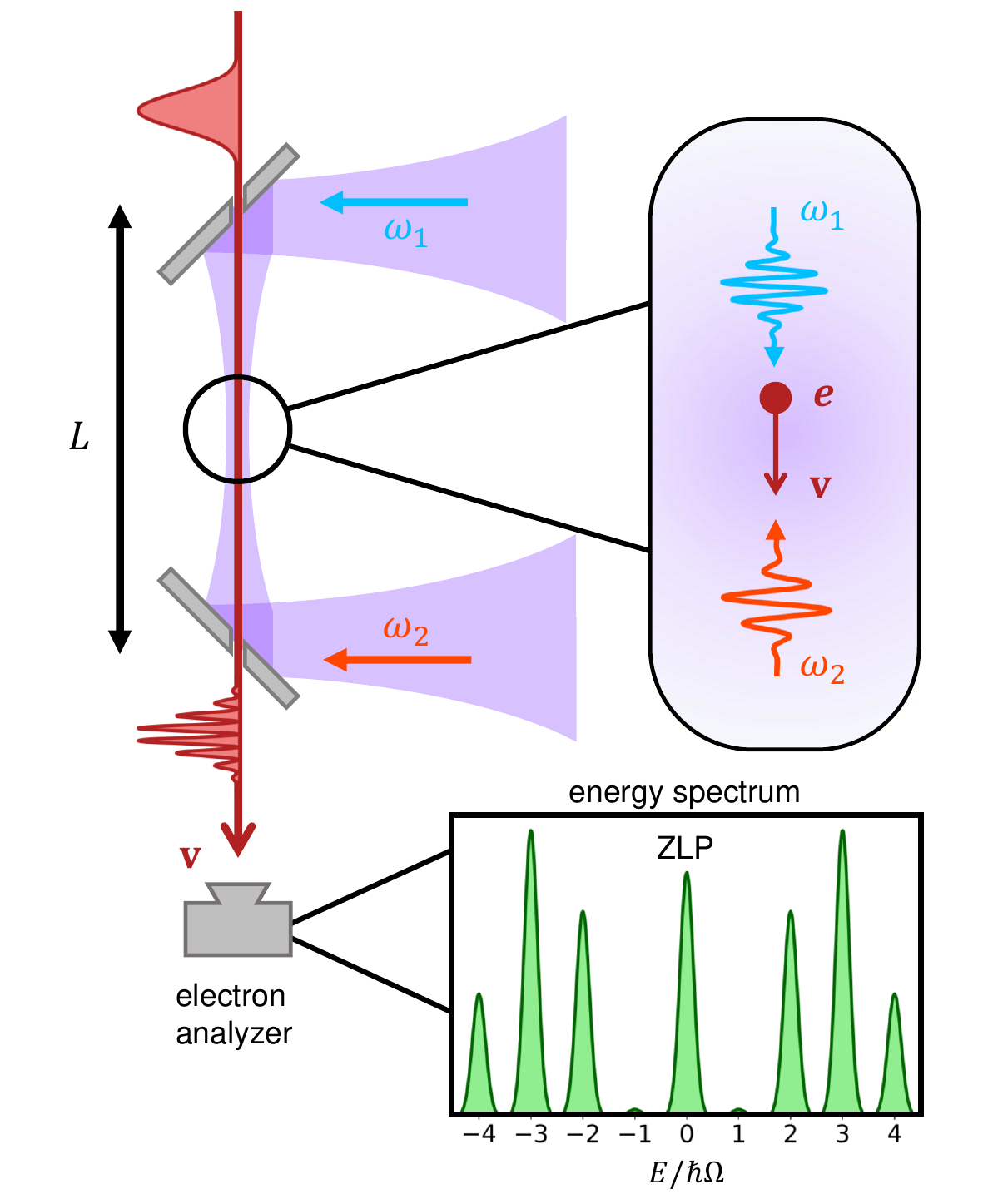}
\caption{Scheme of optical electron modulation in free-space. An electron beam (e-beam) runs along the same axis as two co- and counterpropagating focused laser beams of frequencies $\omega_1$ and $\omega_2$. Through stimulated inelastic Compton scattering, the electron can absorb energy quanta $\hbar\Omega\equiv\hbar(\omega_1-\omega_2)$, producing sidebands in the electron spectrum (bottom inset). Electron propagation after the interaction results in the modulation of the incident electron wavepacket into a train of temporally compressed pulses.}
\label{Fig1}
\end{figure}

In this work, we theoretically demonstrate strong temporal modulation of e-beams in free space under CW operation conditions. We rely on stimulated Compton scattering between two counterpropagating light beams and an e-beam that shares the same optical axis (Fig.~\ref{Fig1}). This configuration enables long millimetric interaction regions over which the amplitude of inelastically scattered electron components can build up to produce a sizeable spectral and temporal modulation. In the nonrecoil approximation (i.e., assuming a constant electron velocity), this scheme leads to electron energy combs completely analogous to those observed in PINEM. However, electron recoil becomes important for long interaction distances, leading to severe corrections in the electron spectra and enabling a dramatically increased degree of temporal compression. We discuss the characteristics of these unexplored physical effects and present simulations showing strong modulation for practical choices of CW electron- and light-beam parameters.

% =========================================================
\section{Results and discussion}

We consider a Gaussian e-beam moving with velocity $v$ along the same optical axis as two CW monochromatic co- and counterpropagating Gaussian laser beams of frequencies $\omega_1$ and $\omega_2$, respectively (Fig.~\ref{Fig1}). Under typical electron-microscope operation conditions, the width of the e-beam remains small compared to the light wavelengths as the electron propagates along the interaction region of length $L$. Therefore, the optical fields are approximately uniform across the transverse directions $(x,y)$ and only depend on the longitudinal coordinate $z$ along the optical axis. This configuration can be realized by reflecting the light beams on mirrors perforated by small holes through which the e-beam is passing. Although single-photon emission or absorption by a moving electron is kinematically forbidden in free space, inelastic Compton scattering with a zero net balance of photon exchanges can conserve energy and momentum. In the present configuration, stimulated processes dominate and result in photon exchanges between the two light beams with a probability proportional to the number of photons in each of them (i.e., the product of the beam powers). Nonstimulated processes in which photons are deflected from the optical axis should contribute negligibly. Energy and momentum are then conserved under the condition
\begin{align}
    \omega_2 = \omega_1\frac{1-v/c}{1+v/c}.
    \label{phasematch}
\end{align}
Individual scattering events produce a change in the electron energy by $\pm\hbar\Omega$, where $\Omega\equiv\omega_1-\omega_2$. The inelastic scattering probability is enhanced for a large interaction length $L$ in the range of a few millimeters.

Under the stated conditions, we can disregard the dependence of the electron wave function on $(x,y)$. The energy spread of the electron is considered to be small compared with its average kinetic energy $E_0=\me c^2(\gamma-1)$, with $\gamma=1/\sqrt{1-v^2/c^2}$, so we write its wave function as $\psi(z,t)=\ee^{\ii(q_0z-E_0t/\hbar)}\phi(z,t)$, where $\hbar q_0=\me v\gamma$ is the central momentum and we introduce a slowly varying envelope $\phi(z,t)$ that satisfies the Schr\"odinger equation \cite{paper368}
\begin{align}
    \bigg[\ii\hbar\big(\partial_t + v\partial_z \big) \! + \! \frac{\hbar^2}{2m\gamma^3} \partial_{zz} \bigg] \phi(z,t) = & \Hint(z,t) \phi(z,t).
    \label{Schrdeq}
\end{align}
The left-hand side of Eq.~(\ref{Schrdeq}) originates from the Taylor expansion of the electron energy as a function of the wave vector, $E_q \approx E_0 + \hbar v (q-q_0) + \hbar^2(q-q_{0})^2/2m\gamma^3$ (i.e., incorporating the lowest-order recoil correction), with the wave vector $q$ substituted by $-\ii\partial_z$. In addition, $\Hint(z,t)=e^2 A^2(z,t)/2mc^2\gamma$ is the minimal-coupling electron--light interaction Hamiltonian, expressed in terms of the classical optical vector potential $A(z,t)$ in a gauge with vanishing scalar potential (see Ref.~\cite{paper368} for a detailed derivation). We only retain $A^2$ terms, which are responsible for the ponderomotive interaction. For swift electrons, linear terms are only contributed by the vector potential component parallel to the velocity, which is negligible in the present configuration, and in addition, those terms cancel upon integration over time because of the noted kinematic mismatch.

We write the total vector potential amplitude associated with the two monochromatic laser beams as
\begin{align}
A(z,t)=\frac{2c\,\Theta(L/2-|z|)}{\sqrt{1+(z/z_0)^2}}\!\sum_{j=1,2}\!\!{\rm Im}\bigg\{\frac{\mathcal{E}_j}{\omega_j}\ee^{\ii(s_jk_jz-\omega_jt)}\bigg\}, \label{Azt}
\end{align}
where $\mathcal{E}_j$ are focal electric field amplitudes, we assume small numerical apertures $\NA_j$ and collinear polarizations in the transverse $x-y$ plane, $z_0=2 c/\omega_j \NA_j^2$ is the Rayleigh range (taken to be the same in both beams for simplicity), the focal spots are both located at $z=0$, and the coefficients $s_1=1$ and $s_2=-1$ indicate co- and counterpropagation relative to the e-beam. The step function limits the interaction region to $-L/2<z<L/2$, and we neglect weak fields scattered by the mirror holes at $z=\pm L/2$, far from the focal region. The power in each beam is $\mathcal{P}_j=c^2z_0|\mathcal{E}_j|^2/2\omega_j$.

We first discuss the interaction in the nonrecoil approximation [i.e., ignoring the $\partial_{zz}$ term in Eq.~(\ref{Schrdeq})]. For a given incident wave function $\phi^{\rm inc}(z,t)$, Eq.~(\ref{Schrdeq}) admits the solution $\phi(z,t)=\phi^{\rm inc}(z,t)\exp\big\{(-\ii/\hbar v)\int_{-L/2}^{L/2} dz' \Hint[z',t+(z'-z)/v]\big\}$ after the interaction has taken place. Inserting Eq.~(\ref{Azt}) into $\Hint$, assuming $L\gg z_0$ (small focal region compared with $L$), and carrying out the $z'$ integral, we find (see \ref{Sec1})
\begin{align}
\phi(z,t)=\phi^{\rm inc}(z,t)\,\ee^{\ii\chi}\sum_{\ell = -\infty}^{\infty} \alpha_\ell\;\ee^{\ii\ell\Omega(z/v-t)},
\label{solutionalphal}
\end{align}
where $\chi$ is a global phase, the sum defines an energy comb with coefficients $\alpha_\ell=J_\ell(2|\beta|)\,\ee^{\ii\ell\arg\{-\beta\}}$, and the interaction strength is fully encapsulated in the parameter
\begin{align}
\beta=\frac{2\pi\ii e^2c}{\hbar m v \gamma}
\;\frac{1+v/c}{1-v/c}
\;\frac{\mathcal{E}_1\mathcal{E}_2^*}{(\NA_1)^2\omega_1^3}.
\label{beta}
\end{align}
We thus obtain electron sidebands corresponding to energy changes $\ell\hbar\Omega$, provided the phase-matching condition in Eq.~(\ref{phasematch}) is satisfied. This equation shows that lower laser frequencies and small NAs produce an increase in the electron--light interaction. Assuming equal powers in both laser beams ($\mathcal{P}_1=\mathcal{P}_2\equiv\mathcal{P}$), we can then plot a universal map for the coupling parameter $|\beta|$, which is proportional to $\mathcal{P}/\hbar\omega_1$ [see Fig.~\ref{Fig1}(a)]. Further setting the electron velocity to $v=c/3$ ($E_0\approx31$~keV kinetic energy), so that $\Omega=\omega_1/2$, we find the sideband probabilities $J_\ell^2(2|\beta|$) shown in Fig.~\ref{Fig1}(b), similar to PINEM, but with the light power instead of the field amplitude in the vertical axis because the interaction Hamiltonian is now $\propto A^2$. In particular, sidebands are effectively populated up to a maximum index $\ell_{\rm max}\sim\sigma_\ell$ determined by the standard deviation $\sigma_\ell=\sqrt{\sum_\ell\ell^2J^2(2|\beta|)}=\sqrt{2}|\beta|$.

% --- Figure 2 --------------------------------------------
\begin{figure}
\centering\includegraphics[width = 0.45\textwidth]{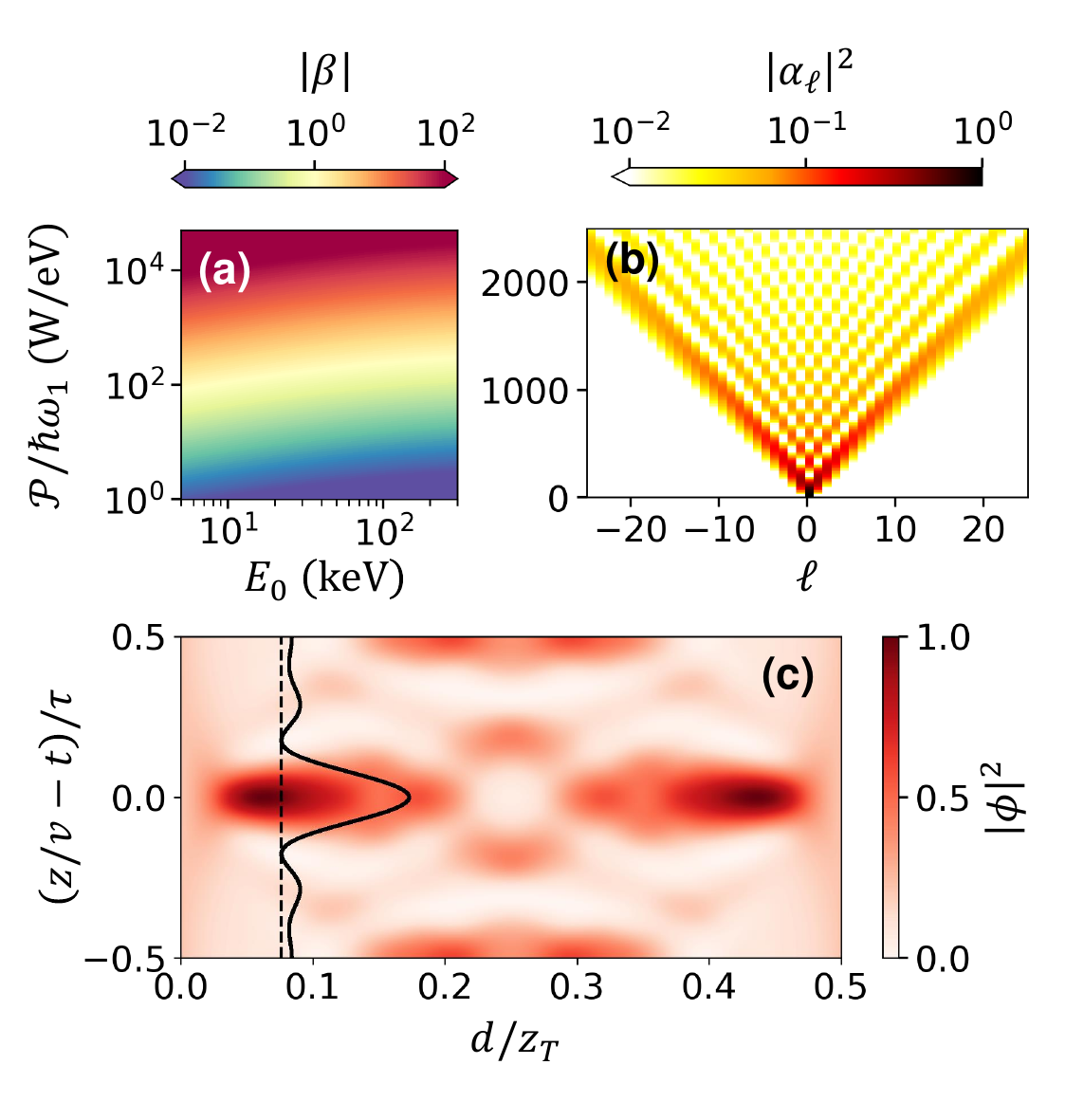}
\caption{Free-space optical electron modulation in the nonrecoil regime. (a)~Electron--light coupling coefficient $|\beta|$ as a function of electron kinetic energy $E_0$ and the ratio of laser-beam power $\mathcal{P}$ to the photon energy in beam 1. (b)~Probability that an electron with incident kinetic energy $E_0=31$~keV ($v=c/3$) gains $(\ell<0)$ or loses $(\ell>0)$ $\ell$ quanta of energy $\hbar\Omega$ during the interaction as a function of $\mathcal{P}/\hbar\omega_1$ (vertical axis). (c)~Evolution of the electron density profile (vertical axis, covering an optical period $\tau=2\pi/\Omega$) for an incident CW electron as a function of propagation distance $d=vt$ (horizontal axis, normalized to the Talbot distance $z_T=4\pi\me v^3\gamma^3/\hbar\Omega^2$) after interaction with $\beta=1$. The curve shows the profile at the optimum compression distance $d/z_T\approx0.076$.}
\label{Fig2}
\end{figure}

From Fig.~\ref{Fig2}(a,c), we conclude that, using mid-infrared light (e.g., $\hbar\omega_1=117$~meV from a CO$_2$ laser), order-unity $\beta$ coefficients and large electron modulation can be achieved for a light power of $\sim10$~W, which would involve an intensity of $\sim10^5$~W/m$^2$ for $L=2$~mm and $\NA_1=0.2$, well below the damage threshold of dielectric mirrors \cite{NFH21}.

The $\partial_{zz}$ term in Eq.~(\ref{Schrdeq}) becomes relevant after a long propagation distance $d=vt$. The solution in Eq.~(\ref{solutionalphal}) is still valid, but the nonrecoil term introduces an additional $\ell$-dependent factor $\ee^{-2\pi\ii\,\ell^2 d/z_T}$ in the coefficients $\alpha_\ell$, where $z_T=4\pi\me v^3\gamma^3/\hbar\Omega^2$ is the so-called Talbot distance \cite{paper360}. The dispersion in the velocity associated with different sidebands causes them to interfere, giving rise to electron wave-function reshaping, as illustrated in Fig.~\ref{Fig2}(c) for $\beta=1$ (a coupling coefficient that could be produced with $\mathcal{P}/\hbar\omega_1\approx170$~kW/eV for $v=c/3$).

From the form of the recoil phase factor, we expect to observe recoil effects in sidebands for which $\ell^2d/z_T$ has unity order (i.e., $d\sim z_T/|\beta|^2$ for the highest sideband order $\ell\sim|\beta|$). An interesting regime is encountered when the size of the interaction region $z_0$ is comparable to or larger than $z_T/|\beta|^2$. Then, the integration of Eq.~(\ref{Schrdeq}) becomes more complex, as we cannot neglect the $\partial_{zz}$ term during the interaction. We formulate an efficient method of solution by concatenating the propagation of the wave function along small $\Delta z\ll z_0$ intervals over which the field envelope function $1/\sqrt{1+(z/z_0)^2}$ [see Eq.~(\ref{Azt})] varies negligibly with $z$. The wave function can then be expanded as $\phi(z,t) = \sum_{\ell_1 \ell_2} \alpha_{\ell_1 \ell_2}(t)\,\ee^{\ii[\omega_1 \ell_1(t-z/c)+\omega_2 \ell_2(t+z/c)]}$ within each interval, and the coefficients $\alpha_{\ell_1 \ell_2}(t)$ are obtained analytically (see \ref{Sec2}), leading to highly convergent results that we corroborate by direct (but lengthy) numerical integration. In what follows, we take the same values of $z_0$ and $\mathcal{P}$ for the two laser beams. Also, the interaction is dominated by resonant processes in which a photon from one of the beams is scattered into a photon in the other beam (see Supplementary Fig.~\ref{FigS1}). Under these conditions, the final electron spectrum only depends on the ratios $\mathcal{P}/\hbar\omega_1$ (number of laser beam photons per unit time), $z_T/z_0$, and $v/c$.

% --- Figure 3 --------------------------------------------
\begin{figure*}
\centering\includegraphics[width=0.95\textwidth]{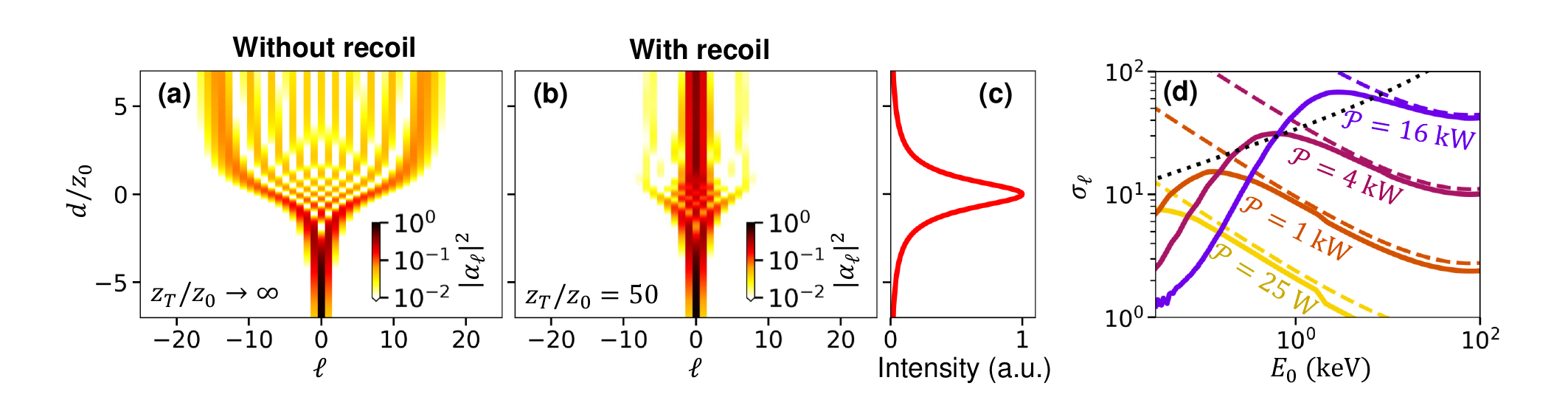}
\caption{Recoil effects in optical electron modulation. (a)~Evolution of the sideband probability in the spectra of 31~keV electrons as a function of propagation distance $d=vt$ inside the interaction region without inclusion of recoil for $\mathcal{P}/\hbar\omega_1=2$~kW/eV. (b)~Same as (a), but including recoil. (c)~Light intensity acting on the electron. (d)~Standard deviation $\sigma_\ell=\sqrt{\sum_\ell \ell^2|\alpha_\ell|^2}$ of the number of sidebands after interaction as a function of electron kinetic energy for $\NA_1=0.2$, $\hbar\omega_1=2$~eV, and different light powers $\mathcal{P}$ (same in both laser beams). We show the nonrecoil result $\sigma_\ell=\sqrt{2}|\beta|$ (dashed curves) and the recoil onset condition $\sigma_\ell=\sqrt{z_T/z_0}$ (dotted curve) for comparison.}
\label{Fig3}
\end{figure*}

% --- Figure 4 --------------------------------------------
\begin{figure*}
\centering\includegraphics[width=1.0\textwidth]{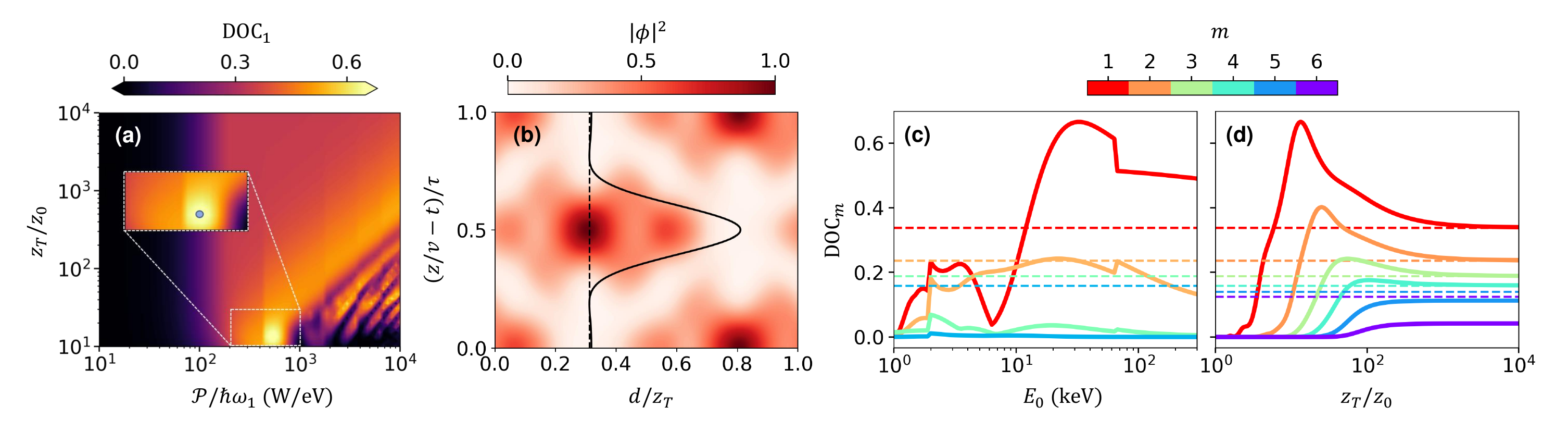}
\caption{Temporal electron compression via ponderomotive interaction. (a)~Maximum degree of coherence (DOC$_1$, order $m=1$) along the evolution of the electron wave function after electron--light interaction as a function of the parameters $\mathcal{P}/\hbar\omega_1$ and $z_T/z_0$ for 31~keV electrons. (b)~Evolution of the electron probability in a representation similar to Fig.~\ref{Fig2}(c), as a function of propagation distance $d=vt$ for 31~keV electrons, $\mathcal{P}/\hbar\omega_1\approx553$~kW/eV, and $z_T/z_0\approx13.3$ [blue dot in (a)]. The curve shows the profile at the position of maximum compression ($d/z_T\approx0.31$). (c)~Maximum DOC$_m$ for different orders $m=1-6$ as a function of electron kinetic energy for the same values of $\mathcal{P}/\hbar\omega_1$ and $z_T/z_0$ as in (b) (solid curves) compared with the nonrecoil result (dashed lines). (d)~Same as (c), but as a function of $z_T/z_0$ for 31~keV electrons and $\mathcal{P}/\hbar\omega_1$ as in (b,c).}
\label{Fig4}
\end{figure*}

In Fig.~\ref{Fig3}, we consider parameters for which recoil becomes important during the interaction region, essentially controlled by the ratio $z_T/z_0$ (see Supplementary Fig.~\ref{FigS2} for more details on the onset of recoil as this parameter is reduced). In particular, in Fig.~\ref{Fig3}(a,b), we plot the sideband probabilitues $|\alpha_\ell|^2$ as a function of propagation distance $d=vt$ without (setting $z_T\rightarrow\infty$) and with (for $z_T=50\,z_0$) inclusion of recoil for an electron velocity $v=c/3$ and $\mathcal{P}/\hbar\omega_1=2$~kW/eV, which imply $|\beta|=12.5$. As a first observation, we note that the bulk of the change in the electron spectrum occurs near the focal region, where the light intensity is maximum [Fig.~\ref{Fig3}(c)]. In addition, recoil produces dramatic effects in the final electron spectrum, with the emergence of a cutoff for a sideband index $\ell\sim7$. To explain this effect, we argue that the spectrum would extend up to $\ell\sim|\beta|$ without recoil, but the recoil phase $-2\pi\ell^2 z_0/z_T\sim1$ becomes important for $\ell\sim\sqrt{z_T/z_0}\approx7$, so dramatic changes in the spectrum occur for sidebands beyond that cutoff. We explore this phenomenon over a wider range of electron energies and different laser powers $\mathcal{P}$ in Fig.~\ref{Fig3}(d) for fixed $\NA_1=0.2$ and $\hbar\omega_1=2$~eV. The results corroborate that the sideband standard deviation $\sigma_\ell$ (solid curves) only differs from the nonrecoil result $\sqrt{2}|\beta|$ (dashed curves) when $\sigma_\ell\gtrsim\sqrt{z_T/z_0}$ (dotted curve).

The recoil causes a departure of the electron velocity from the phase-matching condition in sidebands of large order $\ell$. A similar effect takes place when the initial electron velocity does not fulfill Eq.~(\ref{phasematch}), thus producing a reduction in the effective electron--light interaction that becomes more dramatic as the $\NA$ is reduced and $z_0$ increases (see Supplementary Fig.~\ref{FigS3}; a lower $\NA$ extends the interaction to a larger number of optical cycles, thus better enforcing the phase-matching condition).

As a prominent application of free-electron modulation, we explore temporal compression in Fig.~\ref{Fig4} by calculating the degree of coherence defined as DOC$_m\equiv|I_m/I_0|^2$ with $I_m=\int_{-\pi/\Omega}^{\pi/\Omega} dt'\;|\phi(z=d,d/v+t')|^2\;\ee^{\ii m\Omega t'}$ for integer orders $m$ \cite{paper374}. A maximum compression (point-like electron limit) corresponds to DOC$_m=1$ for all values of $m$. After interaction with light, free propagation reshapes the electron density profile and produces DOC$_m$ values that vary with the propagation distance $d=vt$ (see Supplementary Fig.~\ref{FigS4}). In the absence of recoil, one has DOC$_m=J^2_m[4|\beta|\sin(2\pi md/z_T)]$ \cite{paper373}, so that, at an optimum distance $d$ [see Fig.~\ref{Fig2}(c)], DOC$_m$ reaches a maximum value given by the first maximum of the squared Bessel function $J_m^2$ (e.g., DOC$_1\approx0.34$). However, recoil can have dramatic effects on DOC$_m$ in our ponderomotive modulation scheme. In particular, DOC$_1$ at the optimum $d$ display large variations as a function of $\mathcal{P}/\hbar\omega_1$ and $z_T/z_0$ [Fig.~\ref{Fig4}(a)], and reaches an absolute maximum value of 0.66 for 31~keV electrons with a suitable choice of these parameters. Figure~\ref{Fig4}(b) shows the evolution of the electron density profile with $d$ toward maximal compression at a specific distance for such a combination of parameters (incidentally, the electron spectrum then resembles a Gaussian frequency comb; see Supplementary Fig.~\ref{FigS5}). Similar results are obtained for 200~keV electrons [Supplementary Fig.~\ref{FigS4}(c,d)], showing that this is a robust phenomenon---a conclusion further supported by Fig.~\ref{Fig4}(c), which reveals a region of high compression exceeding the nonrecoil level near 31~keV in the electron-energy dependence of DOC$_m$ for the same choice of $\mathcal{P}/\hbar\omega_1$ and $z_T/z_0$ as in Fig.~\ref{Fig4}(b). Maintaining the kinetic energy at 31~keV and varying $z_T/z_0$ also reveals a range of high compression, followed by convergence to the nonrecoil results for DOC$_m$ as $z_T/z_0$ is increased beyond $\sim10^3$ [Fig.~\ref{Fig4}(d)]. 

% =========================================================
\section{Conclusion}

In conclusion, we have presented a feasible scheme for free-space free-electron optical modulation in the CW regime relying on the increased length of electron--light interaction that is achieved when electron and light beams share the same propagation axis. Stimulated Compton scattering is the underlying mechanism of interaction, requiring phase matching between the electron and light fields over long propagation distances. This condition is broken for electron components that undergo a large number of photon exchanges, thus resulting in electron spectra that differ dramatically from those observed in conventional PINEM experiments. This type of phase mismatch leads to a cutoff in the number of sidebands as well as a large correction in the degree of temporal compression, which is dramatically increased for a suitable combination of laser power and frequency. Our scheme is compatible with existing electron-optics setups, which could simultaneously combine temporal compression and lateral focusing down to the attosecond--sub-{\AA}mstr\"om range. Beyond the emergence of such strong nonrecoil effects, our results support the use of free-space optical modulation as a practical scheme to temporally compress free electrons in free space in a CW regime.

% =========================================================
\section*{ACKNOWLEDGEMENTS} %---ACS---arxiv

This work was supported by the European Research Council (Advanced grant 101141220-QUEFES), the European Commission (Horizon 2020 grant nos. FET-Proactive 101017720-eBEAM and FET-Open 964591-SMART-electron), the Spanish MCINN (PID2020-112625GB-I00 and Severo Ochoa CEX2019-000910-S), the Catalan CERCA Program, and Fundaci\'os Cellex and Mir-Puig.

%\appendix
\renewcommand{\thetable}{A\arabic{table}} %---SI
\renewcommand{\thesection}{Appendix \arabic{section}} %---SI
\setcounter{section}{0}
\begin{widetext}

% =========================================================
\section{Solution of equation~(\ref{Schrdeq}) in the nonrecoil regime}
\label{Sec1}
\renewcommand{\theequation}{A1.\arabic{equation}} %---SI
\setcounter{equation}{0}

Under the conditions stated in the main text for a monochromatic electron traveling with velocity $v$ along the $z$ axis and being exposed to a transverse vector potential $A(z,t)$, we can write the electron wave function as $\psi(z,t)=\ee^{\ii(q_0z-E_0t/\hbar)}\phi(z,t)$, where we separate the fast carrier modulation with momentum $\hbar q_0$ and energy $E_0$ from a slowly varying envelope function $\phi(z,t)$ that satisfies the Schr\"odinger equation
\begin{align}
\bigg[\ii\hbar\big(\partial_t + v\partial_z \big) + \frac{\hbar^2}{2m\gamma^3} \partial_{zz} \bigg] \phi(z,t) = & \Hint(z,t) \phi(z,t)
\label{eqmot}
\end{align}
with $\Hint(z,t)=e^2 A^2(z,t)/2mc^2\gamma$ and $\gamma=1/\sqrt{1-v^2/c^2}$ (Eq.~(\ref{Schrdeq}); see a detailed derivation in Ref.~\cite{paper368}). In the absence of recoil (i.e., neglecting the $\partial_{zz}$ term), this equation admits the solution
\begin{align}
\phi(z,t) =\phi^{\rm inc}(z,t)\; \exp\bigg\{-\frac{\ii}{\hbar v} \int_{-L/2}^{L/2} dz'\;\Hint[z',t+(z'-z)/v]\bigg\},
\label{evol}
\end{align}
where $\phi^{\rm inc}(z,t)$ is the incident electron wave function. Using the vector potential in Eq.~(\ref{Azt}) for $A(z,t)$ and taking $L\gg z_0$, we find the analytical result
\begin{align*}
-\frac{\ii}{\hbar v} \int_{-\infty}^{\infty}  \Hint(z-vt+z',z'/v) dz' = &\ii \pi z_0 \sum_{jj'} \Big[A_{jj'} \ee^{\ii(s_j k_j+ s_{j'} k_{j'})(z-vt)}+A_{jj'}^{*} \ee^{-\ii(s_j k_j+ s_{j'}k_{j'})(z-vt)}\Big] \ee^{-z_0|\Delta_j+\Delta_{j'}|} \\
-&\ii\pi z_0 \sum_{jj'} \Big[B_{jj'} \ee^{\ii(s_jk_j+s_{j'} k_{j'})(z-vt)}+B_{jj'}^{*} \ee^{-\ii(s_j k_j+ s_{j'} k_{j'})(z-vt)}\Big] \ee^{-z_0|\Delta_j-\Delta_{j'}|},
\end{align*}
where we have disregarded the Gouy phase $\ee^{-\ii\arctan(z/z_0)}$ and defined $\Delta_j = \omega_j/v-s_j k_j$, $A_{jj'} = e^2 (\Eb_{j}\cdot\Eb_{j'})/2\hbar mv\gamma \omega_j \omega_{j'}$, and $B_{jj'} = e^2 (\Eb_{j}\cdot\Eb^{*}_{j'})/2\hbar mv\gamma \omega_j \omega_{j'}$. We note that the $A$ and $B$ terms correspond to nonresonant (two-photon absorption or emission) and resonant (Compton scattering) processes, respectively. Assuming the phase-matching condition $\omega_2 = \omega_1\,(1-v/c)/(1+v/c)$ [Eq.~(\ref{phasematch})], we have $\Delta_j-\Delta_{j'} = 0$ for all values of $j$ and $j'$. In addition, if $z_0$ is large compared with the optical period (i.e., assuming that the interaction takes place over many optical cycles of the incident laser fields), we can neglect the $A_{jj'}$ terms and write Eq.~(\ref{evol}) as
\begin{align}
    \phi(z,t) =\phi_{\rm inc}(z,t)\; \ee^{-\ii \chi} \exp\Big\{-\beta \ee^{\ii\Omega(z/v-t)} + \beta^* \ee^{-\ii\Omega(z/v-t)} \Big\},
    \label{shortscale}
\end{align}
where $\Omega \equiv \omega_1-\omega_2$ defines the quantum of frequency exchanges between the light and the electron, while $\chi = 2\pi(B_{11}+B_{22})z_0$ is a global phase that we can ignore in this study. The wave function in Eq.~(\ref{shortscale}) has the same form as in PINEM \cite{paper371}, but with a electron--light coupling parameter now defined as
\begin{align}
    \beta = 2\pi \ii B_{12} z_0 = \frac{2\pi \ii e^2 c}{\hbar m v\gamma} \,  \frac{1+v/c}{1-v/c} \frac{\Eb_{1}\cdot\Eb_{2}^*}{(\NA_1)^2\omega_1^3}.
    \label{coup}
\end{align}
Like in PINEM, using the Jacobi-Anger expansion, we can recast Eq.~(\ref{shortscale}) into Eq.~(\ref{solutionalphal}): $\phi(z,t)=\phi^{\rm inc}(z,t)\,\ee^{\ii\chi}\sum_{\ell = -\infty}^{\infty} \alpha_\ell\;\ee^{\ii\ell\Omega(z/v-t)}$ with $\alpha_\ell=J_\ell(2|\beta|)\ee^{\ii\ell\arg\{-\beta\}}$, showing that an initially monochromatic electron adquires energy sidebands corresponding to energy exchanges $\ell\hbar\Omega$ with a probability $J^2_\ell(2|\beta|)$.

% =========================================================
\section{Solution of equation~(\ref{Schrdeq}) including recoil effects}
\label{Sec2}
\renewcommand{\theequation}{A2.\arabic{equation}} %---SI
\setcounter{equation}{0}

To account for electron recoil, we must solve Eq.~(\ref{eqmot}), including the $\partial_{zz}$ term. We obtain a semi-analytical solution by slicing the interaction region into $N$ segments of length $\Delta z = L/N\ll z_{0}$ centered at positions $\l_n=-L/2+(n+1/2)\Delta z$ with $n=0,\cdots,N-1$, and approximating $1+(z/z_{0})^2\approx 1+(l_n/z_{0})^2$ as a constant within each of them. In practice, we consider $z_0$ to be small compared with $L$ (i.e., the interaction occurs within a small region in between the two mirrors of Fig.~\ref{Fig1}), so we take a large value of $L$ to guarantee a negligible electron--light interaction at $z=\pm L/2$. Using the same vector potential and notation as in Sec.~\ref{Sec1}, Eq.~(\ref{eqmot}) becomes
\begin{align}
\bigg[\ii(\partial_t+v\partial_z)+&\frac{\hbar}{2m\gamma^3}\partial_{zz}\bigg]\,\phi(z,t)\approx \label{eqmotsimp}\\
&-v \sum_{jj} \Big[A_{jj'}^{(n)} \ee^{\ii(s_j k_j+ s_{j'} k_{j'})z-\ii(\omega_j+\omega_{j'})t}+A_{jj'}^{(n)*} \ee^{-\ii(s_jk_j+s_{j'}k_{j'})z+\ii(\omega_j+\omega_{j'})t}\Big] \phi(z,t)  \nonumber \\
&+v \sum_{jj'} \Big[B_{jj'}^{(n)} \ee^{\ii(s_j k_j-s_{j'}k_{j'})z-\ii(\omega_j-\omega_{j'})t}+B_{jj'}^{(n)*} \ee^{-\ii(s_jk_j-s_{j'}k_{j'})z+\ii(\omega_j-\omega_j')t}\Big] \phi(z,t) \nonumber
\end{align}
within segment $n$, where $A_{jj'}^{(n)} = A_{jj'}/[1+(l_n/z_0)^2]$ and $B_{jj'}^{(n)} = B_{jj'}/[1+(l_n/z_0)^2]$ are proportional to $A_{jj'} = e^2 (\Eb_{j}\cdot\Eb_{j'})/2\hbar mv\gamma \omega_j \omega_{j'}$ and $B_{jj'} = e^2 (\Eb_{j}\cdot\Eb^{*}_{j'})/2\hbar mv\gamma \omega_j \omega_{j'}$ (same coefficients as in Sec.~\ref{Sec1}).

At this point, we write the ansatz
\begin{align}
\phi(z,t) = \phi_{\rm inc}(z,t) \sum_{\ell_1,\ell_2 = -\infty}^{\infty} \alpha_{\ell_1 \ell_2}(t) \;\ee^{\ii \ell_1 \omega_1 (t-z/c)+\ii\ell_2 \omega_2 (t+z/c)}
\label{ansatz}
\end{align}
with the initial condition $\alpha_{\ell_1 \ell_2}(-\infty) = \delta_{\ell_1 0} \delta_{\ell_2 0}$ (i.e., an incident monochromatic electron). By introducing Eq.~(\ref{ansatz}) into Eq.~(\ref{eqmotsimp}), we find a system of coupled first-order differential equations for $\alpha_{\ell_1 \ell_2}(t)$. The discretization in $z$ translates into an analogous discretization in time so that we approximate $z\approx l_n$ in the vector-potential profile at times during which the electron is traversing interval $n$ [i.e., for $t_n^-<t<t_n^+$ with $t_n^\pm=(l_n\pm\Delta z/2)/v$]. We find the analytical solution
\begin{align}
\alpha(t_n^+) = \exp\big\{-\ii M^{(n)} \Delta z\big\} \cdot \alpha(t_n^-),
\nonumber
\end{align}
where a matrix notation is used with $(\ell_1\ell_2)$ acting as matrix indices, and the matrix $M^{(n)}$ has time- and position-independent elements
\begin{align*}
    M^{(n)}_{\ell_1 \ell_2,\ell_1' \ell_2'} =& \bigg[ (\ell_1-\ell_2)\omega_1 \bigg(\frac{1}{v}-\frac{1}{c}\bigg) + \frac{\hbar(\omega_1\ell_1 - \omega_2\ell_2)^2}{2mc^2 v\gamma^3} \bigg]\delta_{\ell_1' \ell_1} \delta_{\ell_2' \ell_2} 
    \\
    &+\Big[B_{11}^{(n)}+B_{11}^{(n)*}\Big] \delta_{\ell_1' \ell_1} \delta_{\ell_2' \ell_2} +\Big[B_{22}^{(n)}+B_{22}^{(n)*}\Big] \delta_{\ell_1' \ell_1} \delta_{\ell_2' \ell_2 } +2B_{12}^{(n)} \delta_{\ell_1' (\ell_1+1)} \delta_{\ell_2' (\ell_2-1)}
    \\
    & +2B_{12}^{(n)*} \delta_{\ell_1' (\ell_1-1)} \delta_{\ell_2' (\ell_2+1)} -A_{12}^{(n)} \delta_{\ell_1' (\ell_1+1)} \delta_{\ell_2' (\ell_2+1)} - A_{12}^{(n)*} \delta_{\ell_1' (\ell_1-1)} \delta_{\ell_2' (\ell_2-1)}
    \\
    &-A_{11}^{(n)} \delta_{\ell_1' (\ell_1+2)} \delta_{\ell_2' \ell_2} - A_{11}^{(n)*} \delta_{\ell_1' (\ell_1-2)}\delta_{\ell_2' \ell_2} -A_{22}^{(n)} \delta_{\ell_1' \ell_1} \delta_{\ell_2' (\ell_2+2)} - A_{22}^{(n)*} \delta_{\ell_1' \ell_1} \delta_{\ell_2' (\ell_2-2)}.
\end{align*}
The final wave function is obtained by the concatenation of all intervals, leading to
\begin{align}
\alpha(\infty)=\exp\big\{-\ii M^{(N-1)} \Delta z\big\} \cdots \exp\big\{-\ii M^{(1)} \Delta z\big\}\cdot \exp\big\{-\ii M^{(0)} \Delta z\big\}\cdot\alpha(-\infty).
\label{concatenation}
\end{align}
This approximation fulfills the condition that nonresonant processes (i.e., with $\ell_1+\ell_2\neq0$, so that a net number of photons is exchanged between the light fields and the electron) are negligible [they vanish in the rigorous solution of Eq.~(\ref{eqmot})], as shown in Supplementary Fig.~\ref{FigS1}. The remaining resonant coefficients permit us to obtain the post-interaction coefficients $\alpha_\ell=\alpha_{\ell,-\ell}(\infty)$ in the final wave function
\begin{align}
\phi(z,t)=\phi^{\rm inc}(z,t)\,\ee^{\ii\chi}\sum_{\ell = -\infty}^{\infty} \alpha_\ell\;\ee^{\ii\ell\Omega(z/v-t)} \nonumber
\end{align}
[Eq.~(\ref{solutionalphal})]. We obtain converged solutions by limiting the number of sidebands to $|\ell_1|,|\ell_2|\ll2|\beta|$, where $\beta$ is the nonrecoil coupling coefficient [Eq.~(\ref{coup})].

\end{widetext}

%\bibliography{../../../bibtex/refsU.bib} %---ACS with upper-case title format

\pagebreak

% =========================================================
%\section*{Supplementary Figures}
\renewcommand{\thefigure}{S\arabic{figure}} %---SI
\setcounter{figure}{0}

% --- Figure S1 -------------------------------------------
\begin{figure*}[h]
\centering
\includegraphics[width=0.6\textwidth]{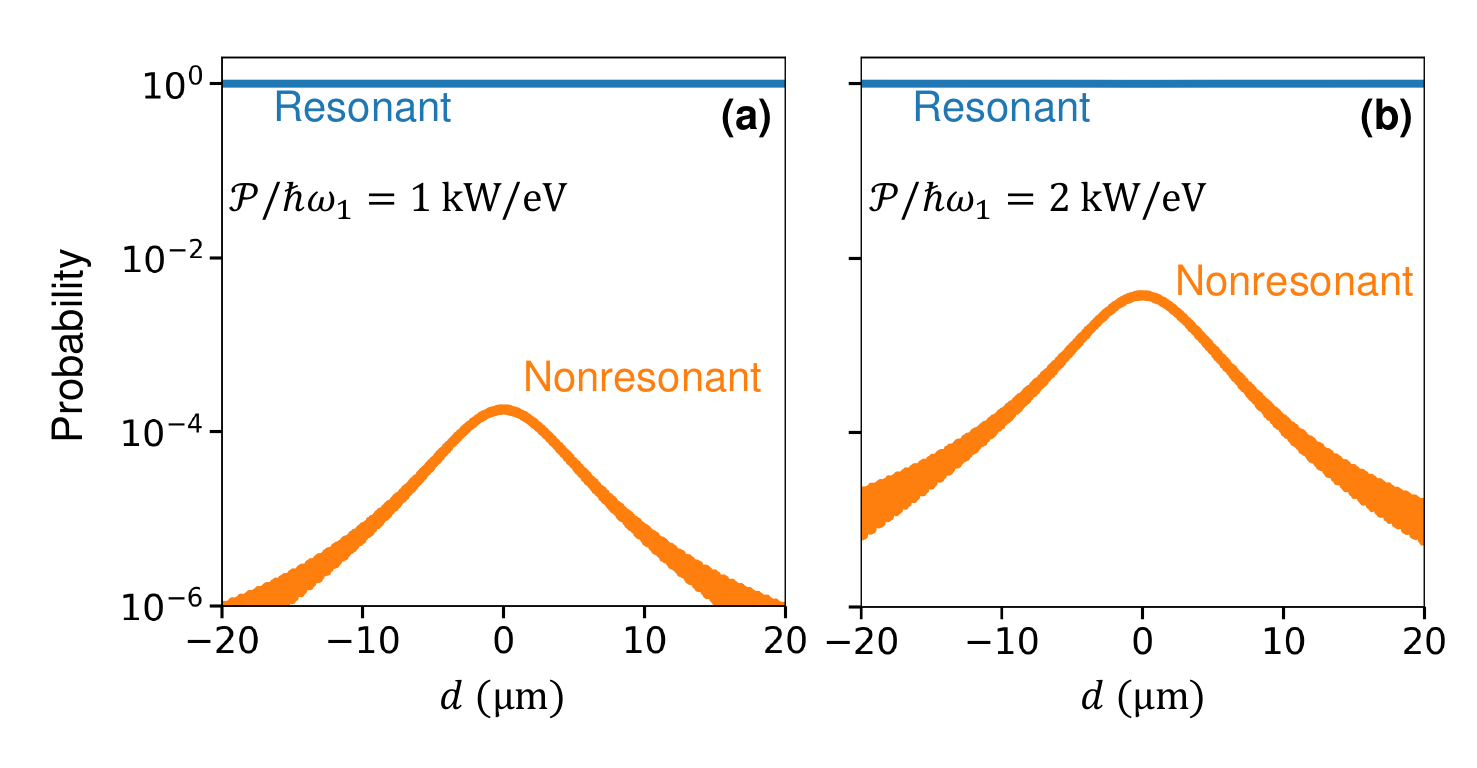}
\caption{Contributions of nonresonant and resonant processes to the interaction probability. We compare the contributions of resonant and nonresonant processes to the total inelastic scattering probability as a function of propagation distance in the interaction region when the ratio $\mathcal{P}/\hbar\omega_1$ between the power and the photon energy in laser beam 1 takes values of 1~kW/eV (a) and 2~kW/eV (b). The electron kinetic energy is 31~keV. Resonant and nonresonant probabilities are computed from $\alpha_{\ell_1\ell_2}(\infty)$ [Eq.~(\ref{concatenation})] as $\sum_\ell|\alpha_{\ell,-\ell}(\infty)|^2$ and $\sum_{\ell_1\neq\ell_2}|\alpha_{\ell_1\ell_2}(\infty)|^2$, respectively}
\label{FigS1}
\end{figure*}

% --- Figure S2 -------------------------------------------
\begin{figure*}[h]
\centering
\includegraphics[width=0.8\textwidth]{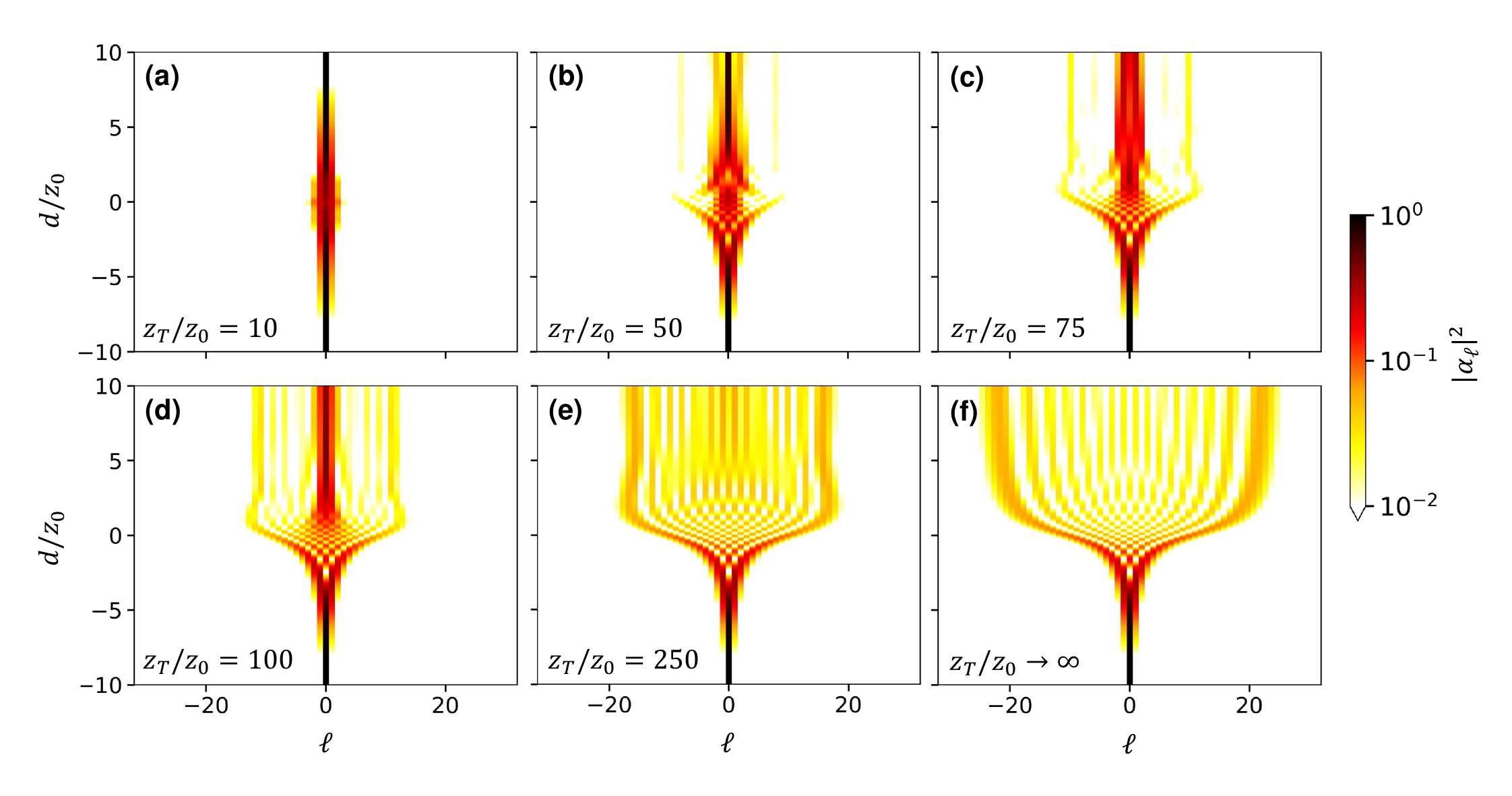}
\caption{Recoil effects in free-space modulation. We explore the onset of recoil effects by plotting the evolution of the sideband probability in the spectra of 31~keV electrons as a function of propagation distance $d=vt$ inside the interaction region for different values of the ratio $z_T/z_0$ (see labels). We set $\mathcal{P}/\hbar\omega_1=4.5$~kW/eV in all calculations.}
\label{FigS2}
\end{figure*}

% --- Figure S3 -------------------------------------------
\begin{figure*}[h]
\centering
\includegraphics[width=\textwidth]{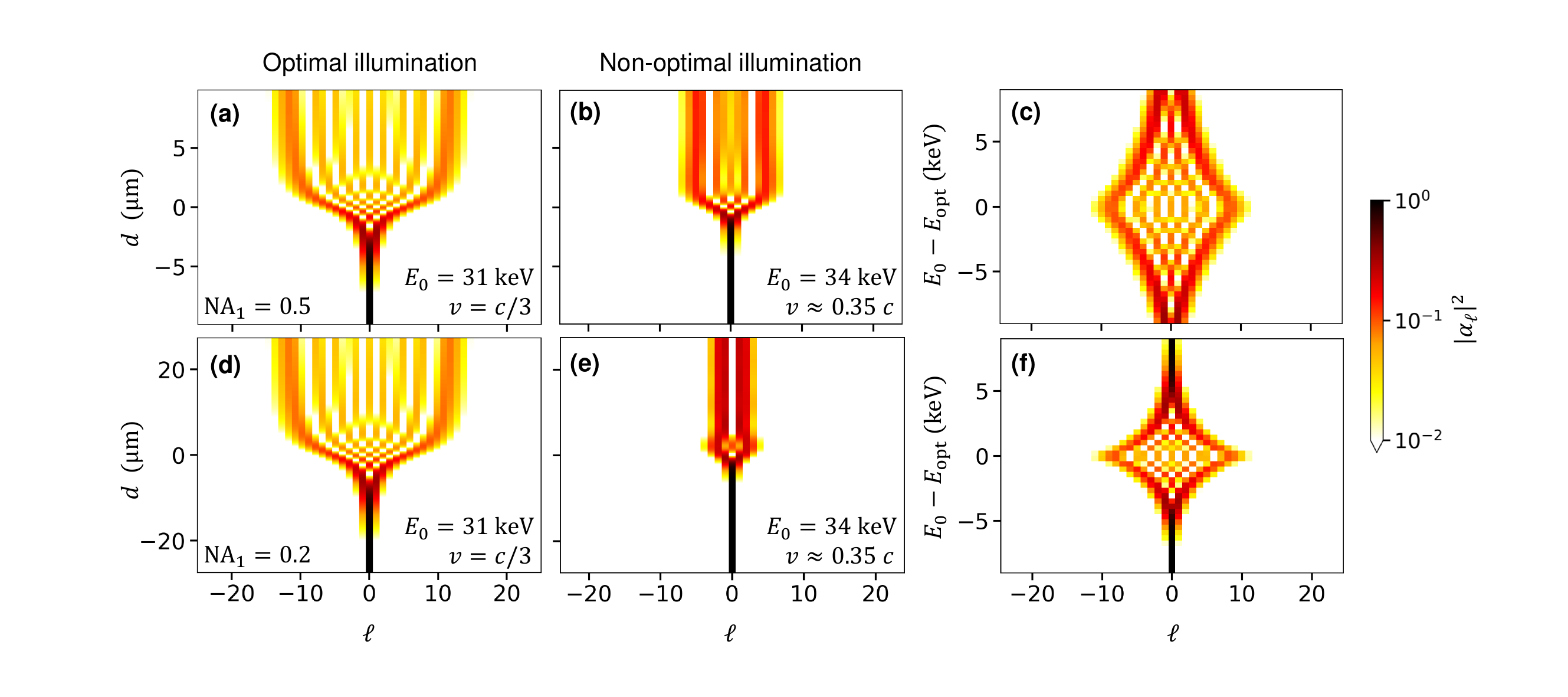}
\caption{Dependence of the electron--light interaction on the phase-matching condition $\omega_2=\omega_1\,(1-v/c)/(1+v/c)$.
(a,d)~Evolution of the sideband probability in the spectra of 31~keV ($v=c/3$) electrons as a function of propagation distance $d=vt$ inside the interaction region for $\hbar\omega_1=2$~eV, $\hbar\omega_2=1$~eV, and $\mathcal{P}=2.5$~kW.
(b,e)~Same as (a,c), but changing the electron energy (velocity) to 34~keV ($v=0.35\,c$) while keeping all other parameters intact.
(c,f)~Electron spectra after the interaction as a function of electron energy $E_0$ relative to the optimum energy $E_{\rm opt}=31$~keV. The numerical aperture $\NA_1$ is 0.5 in (a-c) and 0.2 (d-f).}
\label{FigS3}
\end{figure*}

% --- Figure S4 -------------------------------------------
\begin{figure*}[h]
\centering\includegraphics[width=0.7\textwidth]{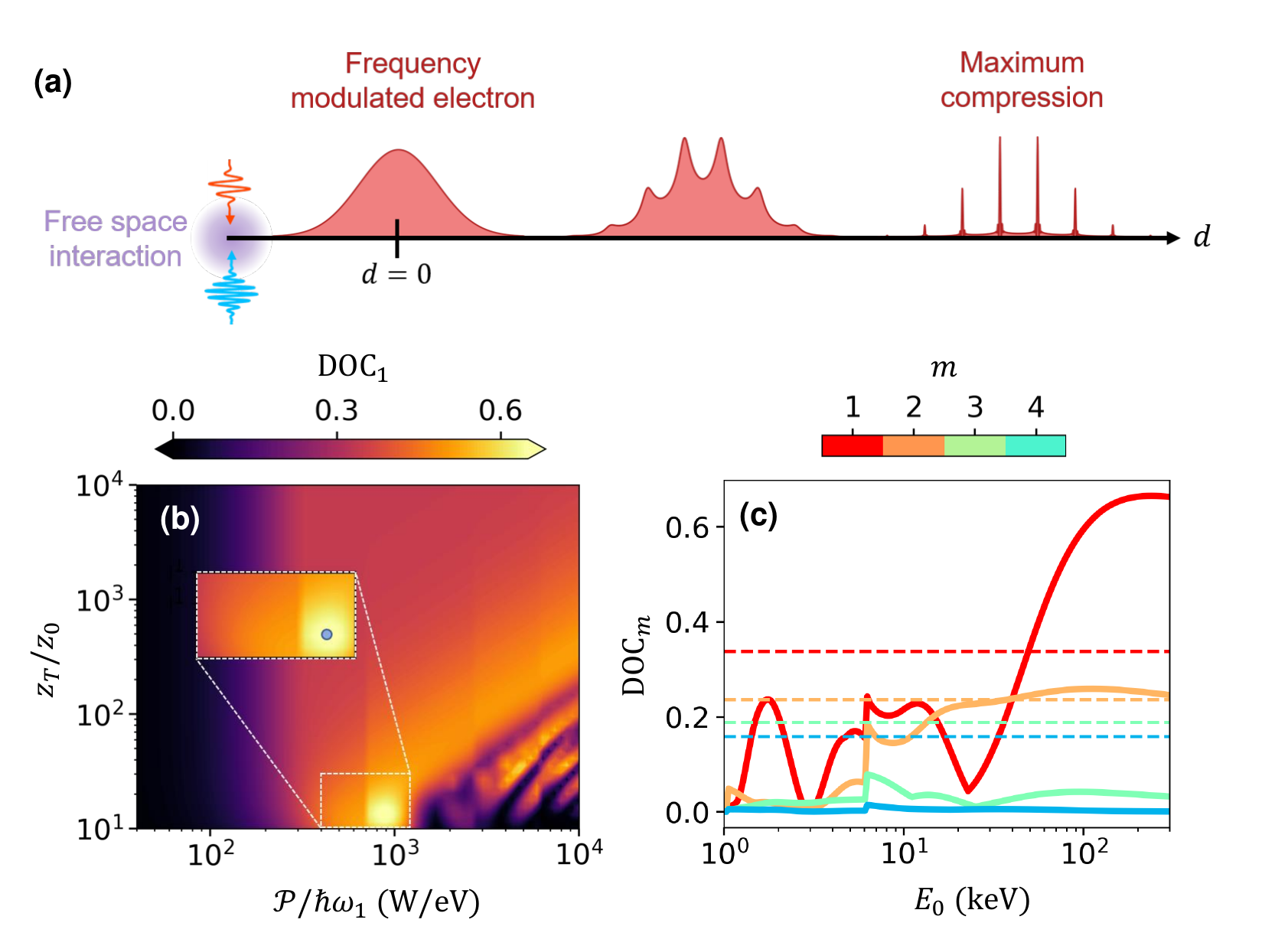}
\caption{Temporal electron compression via ponderomotive interaction. (a)~After the interaction with light, electron components in different energy sidebands move with different velocities and eventually reshape the wave function, becoming maximally compressed at an optimum propagation distance $d=vt$. (b,c)~Same as Fig.~\ref{Fig4}(a,c), but for 200~keV electrons in (a).}
\label{FigS4}
\end{figure*}

% --- Figure S5 -------------------------------------------
\begin{figure*}[h]
\centering\includegraphics[width=0.4\textwidth]{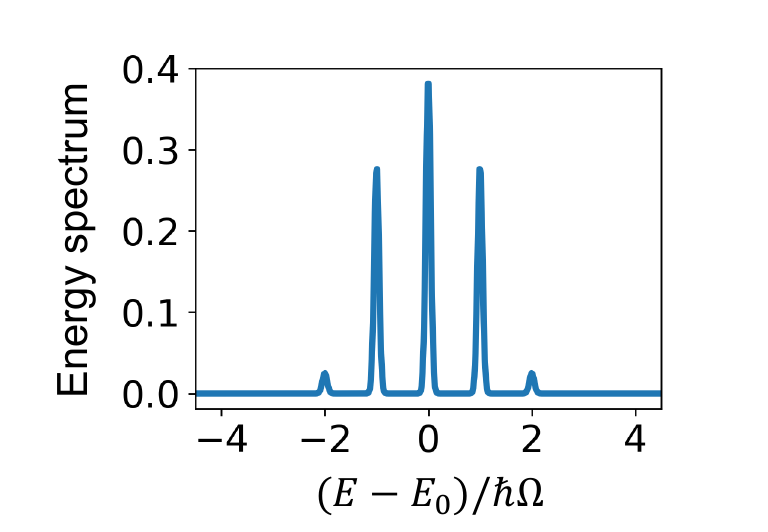}
\caption{Electron spectrum after interaction with light under the conditions leading to optimum compression in Fig.~\ref{Fig4}(b).}
\label{FigS5}
\end{figure*}

\end{document}